\documentstyle[prl,aps,graphicx,multicol]{revtex}
\input psfig.sty
\draft

\begin{document}
\title{Charge ordering in the spinels AlV$_2$O$_4$ and LiV$_2$O$_4$}
\author{Y. Z. Zhang$^1$, P. Fulde$^1$, P. Thalmeier$^2$ and A. Yaresko$^1$}
\address{$^1$Max-Planck-Institut f\"ur Physik komplexer Systeme, N\"othnitzer
Stra\ss e 38 01187 Dresden, Germany\\
$^2$Max-Planck-Institut f\"ur Chemische Physik fester Stoffe,
N\"othnitzer Stra\ss e 40 01187 Dresden, Germany}
\date{\today}
\maketitle

\begin{abstract}
We develop a microscopic theory for the charge ordering (CO) transitions in the spinels
AlV$_2$O$_4$ and LiV$_2$O$_4$ (under pressure). The high degeneracy of CO
states is lifted by a coupling to the rhombohedral lattice
deformations which favors transition to a CO state with inequivalent
V(1) and V(2) sites forming Kagom\'e and trigonal planes respectively.
We construct an extended Hubbard type model including a deformation
potential which is treated in unrestricted Hartree Fock
approximation and describes correctly the observed first-order CO
transition. We also discuss the influence of associated orbital order.  
Furthermore we suggest that due to different band fillings AlV$_2$O$_4$
should remain metallic while LiV$_2$O$_4$ under pressure should become a
semiconductor when charge disproportionation sets in.
\end{abstract}


\begin{multicols}{2}

It was Wigner \cite{Wigner} who first pointed out that electrons form a lattice when the
mutual Coulomb repulsion dominates the kinetic energy gain from
delocalization. He considered an electron gas with a positive uniform
background (jellium). Within that model a lattice will form only when
the electron concentration is very low. Detailed calculations have
determined the critical concentration below which lattice formation or charge
ordering takes place and a value of $r_c\simeq 35a_B$ was found \cite{Imada1}. Here $r_c$
is the critical value of the average distance between electrons which must be
exceeded and $a_B$ is Bohr's radius. The conditions for charge ordering of
electrons are much better when the uniform, positive background is replaced by
a lattice of positive ions. Depending on the overlap of atomic wavefunctions on
neighbouring sites the energy gain due to electron delocalization can be rather
small and a dominance of Coulomb repulsion is more likely. For a more detailed
description see, e.g. \cite{Fulde1}. A good example is Yb$_4$As$_3$ which exhibits
charge order (CO) and for which a microscopic theory was provided
\cite{Fulde2}. Here Yb 4$f$ holes gain a small energy only by delocalizing 
via As 4$p$ hybridization. Another well known example is magnetite Fe$_3$O$_4$,
a spinel structure. Verwey and Haayman \cite{Verwey} found a metal-insulator phase
transition to occur which they attributed to a charge ordering of the Fe$^{2+}$
and Fe$^{3+}$ sites on the pyrochlore sublattice of the spinel lattice
(B-sites). According to Verwey \cite{Verwey} in the charge ordered state the
corner-sharing tetrahedra forming the pyrochlore structure are occupied by two
Fe$^{2+}$ and two Fe$^{3+}$ ions each. This rule, (tetrahedron rule) was most
clearly formulated by Anderson \cite{Anderson} and allows for an exponential
number of different configurations. Recently this rule has been questioned by LDA+U
calculation \cite{Yaresko} which seemingly give good results for the observed charge 
disproportionations \cite{Wright} when the low-temperature structure
is put into the calculations.
 
\begin{figure}[t b]
\begin{center}
\includegraphics[width=7.0cm]{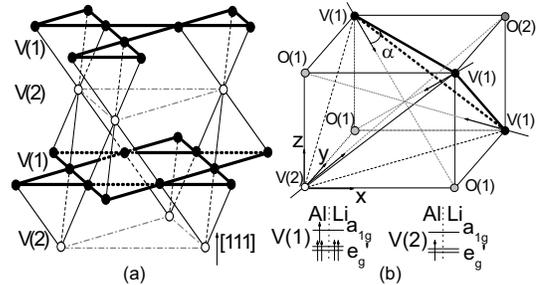}
\end{center}
\vspace{-0.75cm}
\begin{minipage}[t]{8.5cm}
\caption{\label{fig:1}
(a)Sublattice of the V ions. In the presence of a distortion the V(1)
sites form a Kagom\'e and the V(2) sites a triangular lattice. (b)
AlV$_2$O$_4$ and LiV$_2$O$_4$ in the atomic limit. Shown are the
splittings of the $t_{2g}$ orbitals, their occupations, the spin vectors and the
angle $\alpha$ with respect to the [111] plane.}
\end{minipage}
\end{figure}

In this investigation we consider AlV$_2$O$_4$ \cite{Matsuno} and
LiV$_2$O$_4$ \cite{Kondo,Takada}. 
In both systems the $V$ ions form a pyrochlore lattice. While in AlV$_2$O$_4$ a charge order
phase transition at approximately 700 K has been observed
\cite{Matsuno} at ambient pressure, in LiV$_2$O$_4$
electronic charge orders only under hydrostatic pressure of approximately 6 GPa \cite{Takada}. 
In both cases structural changes are associated with charge ordering. 
In AlV$_2$O$_4$ the lattice distortions change the cubic high-temperature phase into a lattice
with alternating layers of Kagom\'e and triangular planes. They are stacked in
[111] direction and formed by nonequivalent V(1) and V(2) sites. The rhombohedral lattice 
deformation causes an increase in the V(1)-O(1) and V(1)-O(2) bond lengths and a decrease 
in the V(2)-O(1) bond length. The changes in the distances V(1)-V(1) 
denoted by $L_1$ and V(1)-V(2) by $L_2$ below the phase transition temperature
T$_{CO}$ define a deformation $\epsilon(T) = (L_2-L_1)/L$. L is the V-V distance 
in the undistorted case. At T = 300 K this value
was found to be $\epsilon(300 K) = 0.016$. Below T$_{CO}$ the average
valence of the V ions changes from +2.5 in the high-temperature phase to
+2.5-$\delta$ at the V(1) sites and to +2.5+3$\delta$ at the V(2) sites. This
is schematically shown in Fig.~\ref{fig:1}. 
\vspace*{-0.3cm}
%%%%%%%%%%%%%%%%%%%%%%%%%%%%%%%%%%%%%%%%%%%%%%%%%%%%%%%%%%%%%%%%%%%%%%%%%%%
%1
%%%%%%%%%%%%%%%%%%%%%%%%%%%%%%%%%%%%%%%%%%%%%%%%%%%%%%%%%%%%%%%%%%%%%%%%%%%
\vspace{-0.3cm}
Also associated with the charge-order transition are a small but steep increase in the resistivity 
and a pronounced decrease of the magnetic susceptibility. In LiV$_2$O$_4$ the lattice distortion 
connected with the charge order under pressure seems to be similar to that of AlV$_2$O$_4$ as powder 
X-ray diffraction indicate \cite{Takada}. However the structural parameters have not yet been determined.
Due to the different filling of the $t_{2g}$ band AlV$_2$O$_4$ seems to remain a semimetal or a small 
gap semiconductor while LiV$_2$O$_4$ is becoming an insulator with a steep increase in the resistivity 
below the phase transition temperature T$_{CO}$.

The first thought is to perform for AlV$_2$O$_4$ and LiV$_2$O$_4$ under pressure LDA+U 
calculation of the type reported for Fe$_3$O$_4$ (see\cite{Yaresko}). We
have done such calculations for AlV$_2$O$_4$.
However, although the LDA+U approach was shown to provide a realistic
description of the electronic structure of magnetic insulators, its
application to paramagnets is less justified because calculations can only
be performed for a magnetically ordered state.
In addition, LDA+U is too crude an approximation to reproduce the low
energy excitation spectrum of strongly correlated metals and, thus, can
hardly be used to study the high temperature metallic phase of AlV$_2$O$_4$
and the change of the electronic structure upon heating through the phase
transition.
For LiV$_2$O$_4$ LDA+U calculations are not yet possible since the CO crystal
structure under pressure is not known.
Therefore we proceed here differently. We 
want to provide a microscopic Hamiltonian in order to identify the physical process which 
results in the observed structural phase transition and the
accompanying charge order while enforcing a proper nonmagnetic state. 
The price are adjustable parameters entering the theory for which no ab initio values are 
available. From standard band-structure calculations it is known that the Fermi energy 
$E_F$ of V 3$d$ electrons lies within the $t_{2g}$ band which is well separated from the 
remaining bands. Therefore we only include t$_{2g}$ electrons in the model defined by
%1 Therefore we shall only include t$_{2g}$ electrons in the model defined by
\begin{eqnarray}
\label{H-1}
&&H = H_0 +H_{int} +H_{e-p}~~,\text{ with} \\
&&H_0 = \sum_{\langle l \mu, l' \mu' \rangle} \sum_{\langle\langle l\mu, l' \mu' \rangle
\rangle \nu \nu' \sigma} t_{\mu \mu'}^{\nu \nu'} \left( l, l' \right) c_{l \mu
\nu \sigma}^+ c_{l' \mu' \nu'\sigma}\nonumber\\  
&&H_{{\rm int}}=\sum_{l\mu }\{(U+2J)\sum_\nu n_{l\mu \nu \uparrow }n_{l\mu
\nu \downarrow }+
U\sum_{\nu >\nu ^{\prime }}n_{l\mu \nu \sigma }n_{l\mu \nu ^{\prime }%
\overline{\sigma }}\nonumber\\
&&+(U-J)\sum_{\nu >\nu ^{\prime }}n_{l\mu \nu \sigma }n_{l\mu
\nu ^{\prime }\sigma }\}+\frac{V}{2}\sum_{\langle l\mu ,l^{\prime }\mu ^{\prime
}\rangle \nu \nu ^{\prime }\sigma \sigma ^{\prime }}n_{l\mu \nu \sigma
}n_{l^{\prime }\mu ^{\prime }\nu ^{\prime }\sigma ^{\prime }}\nonumber\\
&&H_{ep} = \varepsilon \Delta \sum_{l \nu \sigma} \sum_\mu \left( \delta_{\mu,
1} - \frac{1}{3} \left( 1 - \delta_{\mu, 1} \right) \right) n_{l \mu \nu \sigma
} + K \sum_l \varepsilon^2 \nonumber  
\end{eqnarray}
\noindent Here $H_0$ describes the kinetic energy. The electron creation
and annihilation operators are specified by four indices, i.e., for the unit
cell (denoted by $l$), the sublattice ($\mu = 1, 2, 3, 4$), the $t_{2g}$
orbital ($\nu = d_{xy}, d_{yz}, d_{zx}$), and the spin ($\sigma = \uparrow,
\downarrow$). The brackets $\langle ... \rangle$ and $\langle\langle
... \rangle\rangle$ indicate a summation over nearest-neighbour (n.n.) and
next-nearest-neighbour (n.n.n.) sites respectively. The term $H_{\rm int}$
describes the on-site Coulomb and exchange interactions $U$ and $J$ among the
$t_{2g}$ electrons. The last term contains the Coulomb repulsion $V$ of an
electron with those on the six neighbouring sites. Finally 
$H_{ep}$ describes the coupling to lattice distortions. 
The deformation potential is denoted by $\Delta$, it is due to a shift
in the orbital energies of the $V$ sites caused by changes in the
oxygen positions relative to the vanadium positions. 
While the energy shift is positive for the V(1) sites it is negative for the V(2) sites.

The elastic constant $K$ refers to the $c_{44}$ mode and describes the energy due
to the rhombohedral lattice deformation. 
It is reasonable to assume that like in Fe$_3$O$_4$ \cite{Schwenk} and
Yb$_4$As$_3$ \cite{Goto} only the $c_{44}$ mode is strongly coupled with the charge 
disproportionation. We can
give at least approximate values for all parameters except V and
$\Delta$. Their ratio will be fixed by the CO transition temperature
while keeping the constraint V$\ll$U. For the on-site Coulomb- and exchange
integrals we take $U$ = 3.0 eV and $J$ = 1.0 eV which are values commonly used for
vanadium oxides \cite{Fujimori}. Band structure calculations which we have
performed demonstrate that the hopping matrix elements $t_{\mu \mu'}^{\nu
\nu'}$ between different orbitals $\nu \neq \nu'$ are negligible. For
simplicity we first ignore them and omit $\nu, \nu'$. Then 
$t=-t_{\mu\mu'}(l,l')$ when $l$, $l'$ are n.n. and $t'$ for
n.n.n.. We will take into account the orbital dependent
hopping matrix elements coming from a tight-binding fit of LDA
calculations as orbital order is included. Furthermore the $c_{44}$ elastic constant 
is not known for AlV$_2$O$_4$ or LiV$_2$O$_4$,
while computational methods for its ab initio calculation in the case of
materials with strong electronic correlatations are not mature enough. Therefore we use a
representative value $c_{44}^{(0)}/\Omega = 6.1 \cdot 10^{11}$ erg/cm$^3$ 
for AlV$_2$O$_4$ where $\Omega$ is the volume of the cubic unit cell with a lattice 
constant of a = 5.844 $\AA$. This value is close to the experimental value for 
Fe$_3$O$_4$ which has also the spinel structure. This leads to $K \simeq
1.1 \cdot 10^{2} eV$. The deformation potential $\Delta$ is not known
but is commonly of the order of the band width. For convenience we
introduce the dimensionless coupling
constant $\lambda = \Delta^2/Kt$ and lattice distortion
$\delta_L = \epsilon \Delta/t$. From LDA  calculations the bandwidth is
8t=2.7 eV, and therefore a reasonable value is $\lambda
t=\Delta^2/K=$ 1eV. This means $\Delta$ = 10.5eV which is twice the value
of Yb$_4$As$_3$ \cite{Fulde2}.
We begin by considering $H_0$ only. By transforming to Bloch states
$a_{k \xi \nu \sigma}^{+}$ 
%$= \frac{1}{\sqrt{N}} \sum_l c_{l \nu \mu \sigma} e^{{\bf ik} \cdot{\bf R}_l}$, 
we obtain
%5
\begin{eqnarray}
&&H_0=\sum_{{\bf k} \xi \nu \sigma} \epsilon_\xi \left( {\bf k} \right)
a_{{\bf k} \xi \nu \sigma}^{+} a_{{\bf k} \xi \nu \sigma}~~,\text{with}\nonumber\\
&&\epsilon_{1, 2} \left( {\bf k} \right)=2\left( t - t' \right)\\
%\label{epsil12}
&&\epsilon_{3, 4} \left( {\bf k} \right)=-2t \left[ 1 \mp \sqrt{1 + \eta
_{\bf k}} \right] + 2t' \left[ 1 - 2 \eta_{\bf k} \pm 2 \sqrt{1 +
\eta_{\bf k}} \right]\nonumber
\label{epsil34}
\end{eqnarray}
\noindent Here $\eta_{\bf k}=\left(\cos k_x \cos k_y+\cos k_y \cos
k_z+\cos k_z \cos k_x \right)$. With four $V$ ions per unit cell, three
$t_{2g}$ orbitals and two spin directions there are altogether 24 bands, 
of which twelve are dispersionless and degenerate ($\epsilon_{1,2}$). In addition
there are two sixfold 
degenerate dispersive bands present ($\epsilon_{3,4}$). The band structure
has been considered before \cite{Isoda} and we refer to that reference for more
details. The Fermi energy lies in a region of high density of states
(DOS) N(E) which depends sensitively on the ratio $t'/t$. 

Next we discuss the effect of the deformation-potential coupling
described by $H_{ep}$. When it
is sufficiently strong it may lead to charge order. However, there is no gap opening 
but only a sharp decrease of DOS around E$_F$, i.e., the system remains
metallic. When we include the U and V terms in mean-field
approximation, assuming a paramagnetic state, the results remain
qualitatively unchanged. A strong on-site U term suppresses CO while increasing
inter-site V term will induce CO. Again there is no gap opening 
at E$_F$ but a decrease of DOS at the CO transition for
all $t'$. We note that in the present case a n.n.n hopping $t'$
supports CO because it increases N(E$_F$), while 
it prevents CO in the half-filled square lattice \cite{Vekic}. 
Since here $t'<<t$ has little effect on CO we neglect it for simplicity.

In our calculations we assume a homogeneous deformation along [111] direction. 
This yields the following relation between the lattice distortion and charge 
disproportionation
%8
\begin{equation}
\delta_L =\lambda\left( n_2 - n_1 \right)/2
\label{deltalam}
\end{equation}  
\noindent where $n_1=\sum_{\nu\sigma} \langle n_{l 1 \nu \sigma}
\rangle$ is the occupation of a V(2) site while $n_2=n_3=n_4$ with
$n_\mu=\sum_{\nu\sigma}\langle n_{l \mu \nu \sigma}\rangle$ is the
occupation of a V(1) site. 

In any case, the mean-field analysis of the model Hamiltonian (1) leads
to a second order phase
transition as function of temperature. This is at odds with the
experimental observation of a strong first order CO transition. 
We believe that this is the effect of charge correlations caused by
large U and V. To include this effect in a qualitative way, we have
treated the model within an unrestricted Hartree-Fock calculation
by breaking the magnetic symmetry but preserving the constraint of
zero total moment. We assume that the sites $\mu$ have an occupation
$n_\mu=\sum_{\nu\sigma}\langle n_{l\mu\nu\sigma}\rangle$ 
and a magnetization $m_\mu=\sum_{\nu\sigma} \sigma\langle
n_{l\mu\nu\sigma}\rangle$. The spins are assumed to be directed towards the
center of the tetrahedron in the undistorted phase. In the distorted phase
we change the angle $\alpha$ so that the net magnetization remains zero 
(see Fig.~\ref{fig:1} (b)). Because there are two tetrahedra per unit cell translational 
symmetry is maintained. The free energy is 
%9 The free energy is given by
\begin{eqnarray}
&& F  =  -\frac{1}{\beta} \ln \Xi + N_e \mu - \frac{5NU}{12} \sum_\mu n_\mu^2
\nonumber \\ 
&& + \frac{N}{12}\left( U + 4J \right) \sum_\mu m_\mu^2 - NV \sum_\mu \sum_{\mu
\neq \mu'} n_\mu n_{\mu'} + N \frac{\delta ^2}{\lambda}
\label{Fbeta}
\end{eqnarray}
\noindent where $N$ is the total number of unit cell and $N_e$=10 N is the
electron number per unit cell. Furthermore $\Xi$ is the grand canonical
partition function
%10
\begin{equation}
\ln \Xi = 3 \sum_{{\bf k}\xi \sigma } \ln \left[ 1 + e^{-\beta \left(
\epsilon_\xi^\sigma \left( {\bf k} \right) - \overline{\mu} \right) }\right]
\label{lnXi}
\end{equation}
\noindent with $\beta = 1/k_BT$. The chemical potential
$\overline{\mu}$ is adjusted to yield the correct filling, i.e., $N_e = \frac{1}{\beta}
\frac{\partial}{\partial{\overline{\mu}}} ln \Xi$. The energy bands
$\epsilon^\sigma_\xi ({\bf k})$ depend now on spin $\sigma$ for which the [111]
direction is chosen as quantization axis. They are computed by diagonalizing
the Hamiltonian (\ref{H-1}) in mean-field approximation and by
self-consistent determination of $n_\mu$ and $m_\mu$ from $\left(
\partial F/\partial n_\mu\right)_{\overline{\mu}} = 0$ and $\left(\partial
F/\partial m_\mu \right)_{\overline{\mu}} = 0$, respectively at various
temperatures $T$.  

With the magnetization pattern described above and shown in
Fig.~\ref{fig:1} (b) we must transform the nearest-neighbour hopping matrix
elements so that the different spin directions are accounted for. We label the
electronic spin functions of sublattice 1 by $|\sigma_1 \rangle$ and
$|\overline{\sigma}_1 \rangle$ and those referring to sublattice 2, 3, and 4 by
$|\sigma_\mu \rangle$ and $|\overline{\sigma}_\mu \rangle$. We then
apply the transformation to a local spin quantization axis

%11
\begin{equation}
\left( \begin{array}{l}
\mid \sigma_\mu \rangle\\
\mid\overline{\sigma}_\mu \rangle
\end{array} \right) 
= \left( \begin{array}{lr}
\cos \frac{\theta}{2} & -e^{i\gamma_\mu} \sin \frac{\theta}{2}\\
e^{-i\gamma_\mu} \sin \frac{\theta}{2} & \cos \frac{\theta}{2}
\end{array} \right) 
\left( \begin{array}{l}
\mid \sigma_1 \rangle\\
\mid\overline{\sigma}_1 \rangle
\end{array} \right)
\label{matrsigmu}
\end{equation}

\noindent where $\theta=\frac{\pi}{2}+\alpha$ and $\gamma_\mu=0,
\frac{2\pi}{3}, \frac{4\pi}{3}$ for $\mu$=2, 3, 4. All the interaction terms as
well as the electron-phonon coupling term remain invariant under this
transformation. The angle $\alpha$ is adjusted such that $\sum_\mu
m_\mu$=0 for all distortions. 
\vspace*{-0.4cm}
%%%%%%%%%%%%%%%%%%%%%%%%%%%%%%%%%%%%%%%%%%%%%%%%%%%%%%%%%%%%%%%%%%%%%%%%%%%
%3
\begin{figure}[t b]
\begin{center}
\includegraphics[width=6.0cm]{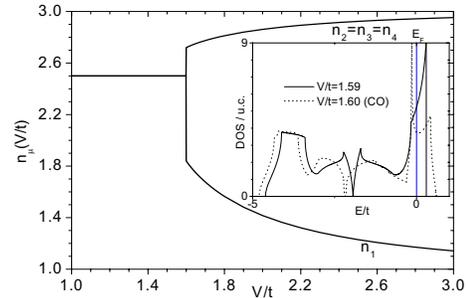}
\end{center}
\vspace{-0.75cm}
\begin{minipage}[t]{8.5cm}
\caption{\label{fig:2}
Charge disproportionation as function of V/t. The $n_\mu$ denote the
occupation numbers of the four sites of a tetrahedron. In the inset the changes in
the density of states are shown when V/t is just below and above the critical
value at which charge ordering sets in. At the critical V/t=1.67
n$_1$=2.5+3$\delta$ and n$_i$=2.5-$\delta$ (i=2-4) with a charge
disproportionation $\delta\simeq$ 0.25. Here U=3.0eV, J=1.0eV,
$\lambda t$=1.0eV and 8t=2.7eV.}
\end{minipage}
\end{figure}
%%%%%%%%%%%%%%%%%%%%%%%%%%%%%%%%%%%%%%%%%%%%%%%%%%%%%%%%%%%%%%%%%%%%%%%%%%%
\vspace{-0.3cm}
The result for the charge disproportionation at $T$=0 as
function of the electronic nearest-neighbour Coulomb repulsion $V$ is shown in
Fig.~\ref{fig:2}. As usual the charge disproportionation is larger (by a factor 
of 2.5 here) than the one obtained from a valence bond analysis of the
distorted structure. This is due to the simplified Hamiltonian which 
cannot describe the screening effects of non-d electrons. The same observation 
was made for Yb$_3$As$_4$ \cite{Fulde2}.
Also shown as an inset is the change in the DOS for a $V$
value just below and above the critical value at which charge order does
occur. Apparently the DOS changes considerably near $E_F$ due to charge
order. This may explain the drop in the susceptibility and the small but steep
increase in the resistivity in AlV$_2$O$_4$ when charge order sets in. The self-consistent
field calculations were done for different temperatures. The phase transition
to the distorted phase is found to be of first order. Fixing the value
V/t=1.67 leads to the calculated transition temperature is $T_c$=660 K
in reasonable agreement with the experimental value
of $T_c^{\rm exp}$=700 K. The phase diagram is found
to be very simple. For $U$=3 eV, $J$=1.0 eV and $8t$=2.7 eV we find a phase
boundary line of the form $\lambda t$+3V=7.8$t$. For all $\lambda t$ values
above that line the trigonal, i.e., distorted phase is found while for values
below that line the phase remains cubic. 

It is obvious that with a lattice distortion an associated 
orbital order will occur \cite{Imada2}. The point symmetry in the charge ordered 
state is $D_{3d}$ and lifts partially the 3-fold degeneracy of the $t_{2g}$
states. This results in one $a_{1g}$ orbital and 2-fold degenerate $e'_g$
orbitals. Experiments show in the distorted
phase an elongation in [111] direction and a corresponding contraction perpendicular to that
direction. Therefore the orbital energy of the $a_{1g}$ state is expected to be
higher than that of the $e'_g$ states.

From the behavior in the atomic limit (see Fig. 1 (b)) we expect that
the system remains gapless 
in the charge ordered state even when orbital order occurs. This is
indeed what we find. We have adopted hopping matrix elements which reproduce
the LDA band structure. They are $t_0=t_{14}^{xy,xy}$=-0.525 eV and
$t_1=t_{12}^{xy,xy}=t_{13}^{xy, xy}$= 0.152 
eV. The upmost band remains almost flat. By solving self-consistently for the
mean-field parameters $n_{\mu\nu}$ and $m_{\mu\nu}$ under the constraint that
the occupation numbers of the three V(1) sites remain equal we find that
$n_2^{xy}=n_2^{zx} > n_2^{yz}, n_3^{xy}=n_3^{yz}>n_3^{zx}$ and
$n_4^{yz}=n_4^{zx}>n_4^{xy}$. For simplicity we have used here the notation of
the $t_{2g}$ basis. It shows that the orbital orders on sublattices 2, 3 and 4
are perpendicular to each other. At the V(2) sites there is no orbital ordering
in the charge ordered phase. Since the system remains gapless and the orbital order 
does not change the effects induced by charge order and magnetic order, i.e., a 
sharp decrease of DOS around the Fermi surface and a first-order  
phase transition respectively, its role is insignificant at 5/12-filling (AlV$_2$O$_4$). 

\begin{figure}[t b]
\begin{center}
\includegraphics[width=5.0cm]{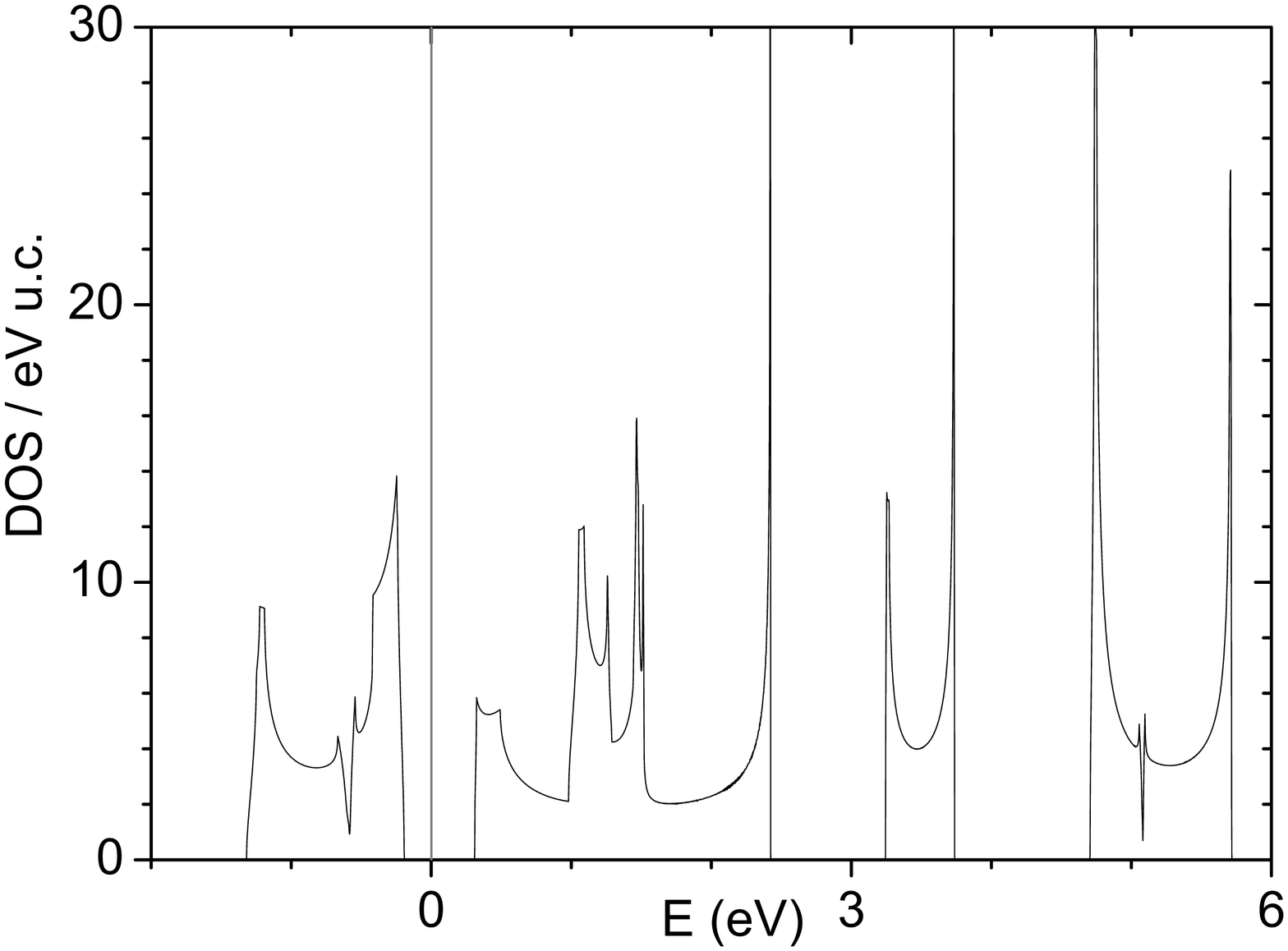}
\end{center}
\vspace{-0.75cm}
\begin{minipage}[t]{8.5cm}
\caption{\label{fig:3}
Density of states of LiV$_2$O$_4$ in the distorted phase. It is seen that
the system is a semiconductor with a gap $E_g$=0.49eV. Here U=3.0eV,
J=1.0eV, V=0.7eV, $\Delta^2/K$=1.0eV, $t_0$=-0.53eV, $t_1$=0.15eV}
\end{minipage}
\end{figure}

Now we compare the above findings with the charge order observed
in LiV$_2$O$_4$ under pressure. Here the average $d$ electron number per $V$ site
is 1.5. The atomic limit shown in Fig.~\ref{fig:1} (b) suggests a gap and
therefore insulating behavior in the charge ordered phase. This is also what is
found when the Hamiltonian (\ref{H-1}) is treated in mean-field approximation
for 1/4 filling as required for LiV$_2$O$_4$ instead of 5/12. The
essential role for orbital order and gap formation is played by the
different 3d-occupation of LiV$_2$O$_4$ therefore we use similar values
for hopping terms and interaction strengths are used as for
AlV$_2$O$_4$. The DOS at $T$=0 is shown in Fig.~\ref{fig:3}. The opening of a gap is
in agreement with the behavior of $\rho(T)$ as found in Ref. \cite{Takada}.
\vspace*{-0.3cm}
%%%%%%%%%%%%%%%%%%%%%%%%%%%%%%%%%%%%%%%%%%%%%%%%%%%%%%%%%%%%%%%%%%%%%%%%%%%
%4
%%%%%%%%%%%%%%%%%%%%%%%%%%%%%%%%%%%%%%%%%%%%%%%%%%%%%%%%%%%%%%%%%%%%%%%%%%%
\vspace{-0.3cm}
In summary, we have provided a microscopic model for the observed charge order
in AlV$_2$O$_4$ and LiV$_2$O$_4$ under pressure which is mainly driven by
a deformation potential coupling to t$_{2g}$ states. The model
accounts correctly for the observed first order CO transition. We have
also shown that due to a difference in the filling factors AlV$_2$O$_4$
should remain metallic in the charge ordered phase while LiV$_2$O$_4$
should become an insulator or semiconductor.

We thank Drs.  G. Venketeswara Pai, T. Maitra, E. Runge and Y. Zhou for helpful discussions.

\end{multicols}

\end{document}